\documentclass[reprint,amsmath,amssymb,aps]{revtex4-2}

\usepackage{graphicx}
\usepackage{dcolumn}
\usepackage{bm}
\usepackage{url}
\usepackage{here}
\begin{document}
\title{
Quantum liquid states of spin solitons in a ferroelectric spin-Peierls state
}

\author{Shusaku Imajo$^{1}$}
\email{imajo@issp.u-tokyo.ac.jp}
\author{Atsushi Miyake$^{1,2}$}
\author{Ryosuke Kurihara$^{1,3}$}
\author{Masashi Tokunaga$^{1}$}
\author{Koichi Kindo$^{1}$}
\author{Sachio Horiuchi$^{4}$}
\author{Fumitaka Kagawa$^{5,6}$}
\affiliation{
$^1$Institute for Solid State Physics, University of Tokyo, Kashiwa, Chiba 277-8581, Japan\\
$^2$Institute for Materials Research, Tohoku University, Oarai, Ibaraki 311-1313, Japan
$^3$Department of Physics, Faculty of Science and Technology, Tokyo University of Science, Noda, Chiba 278-8510, Japan\\
$^4$Research Institute of Advanced Electronics and Photonics (RIAEP), National Institute of Advanced Industrial Science and Technology (AIST), Tsukuba, Ibaraki 305-8565, Japan\\
$^5$Department of Physics, Tokyo Institute of Technology, Meguro, Tokyo 152-8551, Japan\\
$^6$RIKEN Center for Emergent Matter Science (CEMS), Wako, Saitama 351-0198, Japan
}

\date{\today}

\begin{abstract}
In this study, we performed high-magnetic-field magnetization, dielectric, and ultrasound measurements on an organic salt showing a ferroelectric spin-Peierls (FSP) state, which is in close proximity to a quantum critical point.
In contrast to the sparsely distributed gas-like spin solitons typically observed in conventional spin-Peierls (SP) states, the FSP state exhibits dense liquid-like spin solitons resulting from strong quantum fluctuations, even at low fields.
Nevertheless, akin to conventional SP systems, a magnetic-field-induced transition is observed in the FSP state.
In conventional high-field SP states, an emergent wave vector results in the formation of a spin-soliton lattice.
However, in the present high-field FSP state, the strong quantum fluctuations preclude the formation of such a soliton lattice, causing the dense solitons to remain in a quantum-mechanically melted state.
This observation implies the realization of a quantum liquid--liquid transition of topological particles carrying spin and charge in a ferroelectric insulator.
\end{abstract}

\maketitle
\begin{figure}
\centering\includegraphics[width=0.8\hsize]{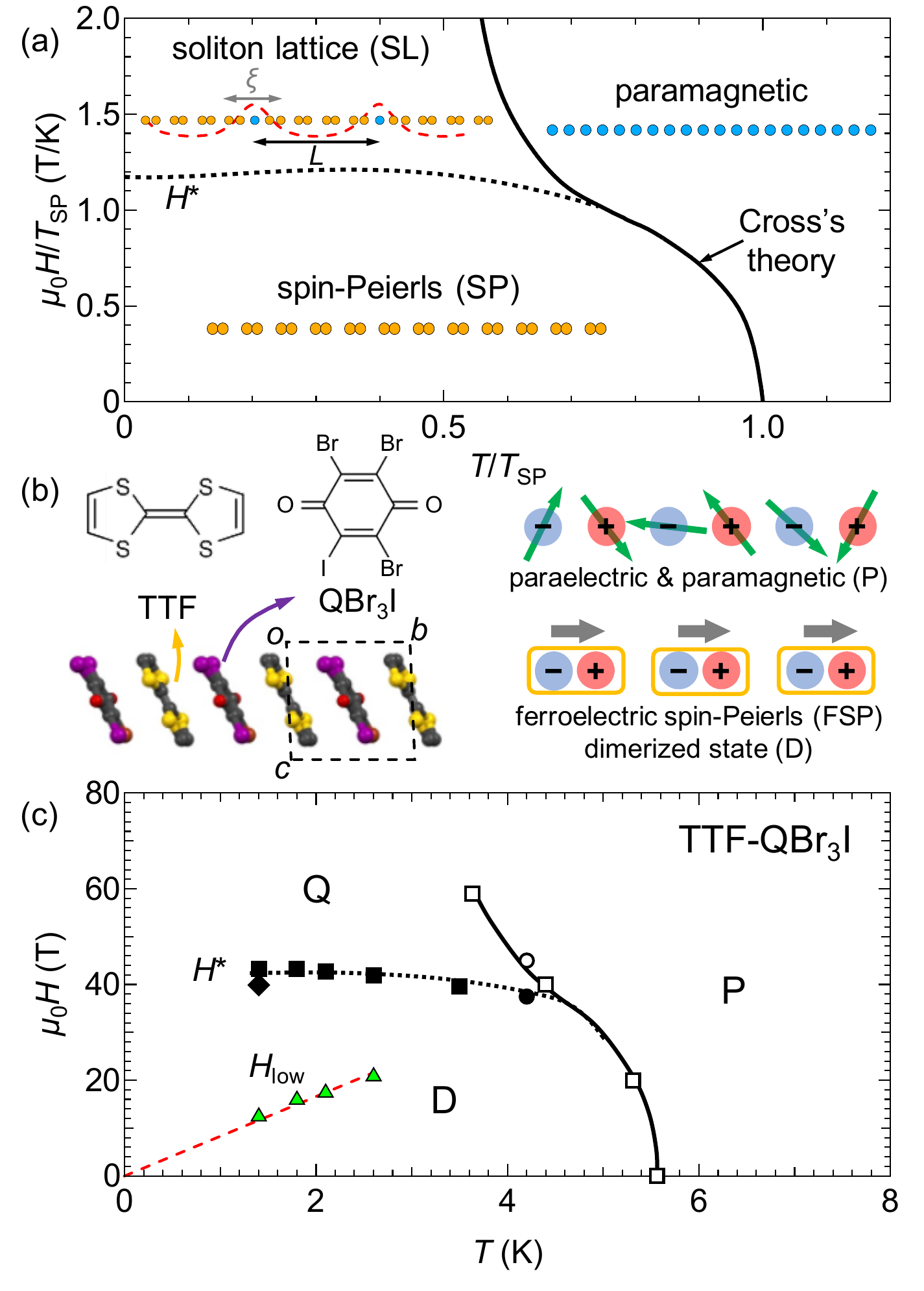}
\caption{
(a) Reduced field-temperature phase diagram of a conventional spin-Peierls (SP) state.
Solid curve shows theoretical calculation of field dependence of the SP transition temperature based on Cross's theory\cite{5,19}.
Dotted curve represents SP-SL phase boundary determined experimentally\cite{12,14,26}.
Insets show schematic illustration of spin sites in a one-dimensional chain.
Blue circles signify non-dimerized sites, whereas orange circles signify dimerized sites with finite lattice displacement.
Red dashed curve of the SL state indicates site-dependent average spin polarization.
(b) Schematic illustration of one-dimensional chain of TTF and QBr$_3$I molecules for paraelectric/paramagnetic (P) state and dimerized (D) state of ferroelectric spin-Peierls (FSP) state.
Green arrows represent spins, while the site colors indicate the ion type (red: cation, blue: anion).
For the D state, gray arrows show the electric polarization of the dimers.
(c) $H$-$T$ phase diagram of TTF-QBr$_3$I.
Here, the high-field state is referred to as the Q state.
Squares and triangles denote anomalies in the temperature/field dependence of dielectric permittivity.
Circles and diamond signify the anomaly fields observed in ultrasound and magnetization measurements, respectively.
Curves on the symbols serve as visual guides.
}
\label{fig1}
\end{figure}

The application of a high magnetic field to a quantum state can induce modifications in its ground state, leading to the emergence of nontrivial periodic structures.
Examples of such phenomena include the Fulde--Ferrell--Larkin--Ovchinnikov (FFLO) state in singlet-pairing superconductivity\cite{1,2} and field-induced density wave (DW) states in low-dimensional metals\cite{3,4}.
In a one-dimensional (1D) quantum magnet with a spin of S = 1/2, its 1D instability gives rise to a long-range ordered state known as the spin-Peierls (SP) state at low temperatures, attributable to the coupling between spins and lattice\cite{5,6}.
In the SP state, an energy gain is obtained by the formation of a spin gap at the expense of the energy cost associated with lattice dimerization.
Upon the application of a magnetic field to the SP state, the spin gap gradually diminishes, accompanied by a decrease in a transition temperature, $T_{\rm SP}$($H$), as shown in Fig.~\ref{fig1}a.
However, this simplistic understanding breaks down when the field reaches a critical value, $H^{\ast}$, where the spin gap and Zeeman energy become comparable.
Above $H^{\ast}$, a spatially modulated state becomes more stable to utilize the Zeeman energy.
The periodic modulation in the high-field SP state induces polarized spin solitons at the domain boundaries, where the order parameter changes sign, resulting in the formation of a spin-soliton lattice (SL) with the corresponding wave vector, as depicted in the inset of Fig.~\ref{fig1}a\cite{9,10,12,13,14,15,16,17}.
The high-field SL phase has been examined in both theoretical\cite{5,6,9,10,14,19,20,21,22,23} and experimental\cite{12,13,14,15,16,17,24,25,26,27,28,29,30} studies, providing valuable insights into its magnetic properties and phase diagram, as shown in Fig.~\ref{fig1}a.

When the SP state is coupled with other degrees of freedom, novel emergent phenomena can occur.
As a prime example, the 1D organic salt TTF-QBr$_4$ (TTF=tetrathiafulvalene, QBr$_4$=$p$-bromanil) exhibits a ferroelectric spin-Peierls (FSP) transition from a paramagnetic and paraelectric (P) state at 53~K\cite{31,32,33}.
The FSP state is realized by the dimerization of magnetic cations and anions in 1D chains, as shown in Fig.~\ref{fig1}c.
The dimer-singlet state, which is the origin of the ferroelectric polarization, is suppressed in magnetic fields.
Consequently, a magnetic-field-controllable ferroelectricity emerges as a result of the coupling with the electric degree of freedom\cite{31}.
However, for TTF-QBr$_4$, the high-field phase has yet to be observed due to the presence of a large spin gap, which restricts experimental access to the high-field regime.
Recently, we reported that the analogous salt TTF-QBr$_3$I (QBr$_3$I=2-iodo-3,5,6-tri-bromo-$p$-benzoquinone) also exhibits a transition from the P state to the FSP state below 5.6~K\cite{34}.
The substitution of the acceptor molecule from QBr$_4$ to QBr$_3$I works as a negative chemical pressure\cite{34,35} and shifts the FSP state to near a quantum critical point.
The developed quantum fluctuations render the topological spin solitons mobile, even in low-temperature regions\cite{34}.
Therefore, it is expected that strong quantum fluctuations and coupling with ferroelectricity potentially induce a nontrivial high-field state and unique soliton physics.

In this study, we investigated TTF-QBr$_3$I using various high-field measurements up to 60~T, and successfully established a field-temperature phase diagram of the FSP state, as shown in Fig.~\ref{fig1}c.
The results show that the dimerized (D) state shows a transition to the high-field phase (hereinafter, this high-field FSP state is referred to as the Q state) above 40-45~T (black symbols).
We find that the field-temperature phase diagram of the FSP state is very similar to those of superconductivity and DW states\cite{4,7,8}.
Our present findings facilitate the understanding of the effect of the magnetic field not only on the SP state coupled with ferroelectricity but also other Zeeman-energy-driven phenomena.
Additionally, we discuss the realization of the quantum liquid--quantum liquid transition of spin solitons induced by a magnetic field.
Considering the spin soliton as a topological particle, the exploration of its analogies with diverse topologically protected magnetic particles, such as chiral solitons and skyrmions\cite{30p5,30p6}, constitutes an intriguing avenue of inquiry.
While topological particles in real space typically lose their dynamics at low temperatures, and their quantum liquid states have not been well examined, the findings presented herein are poised to catalyze future investigations into the quantum effects on topological particles.

Experimental details are shown in Supplementary Materials\cite{Supp}.
Figure~\ref{fig2}a shows the magnetization curves of TTF-QBr$_3$I at 1.4~K.
\begin{figure*}
\centering\includegraphics[width=0.9\hsize]{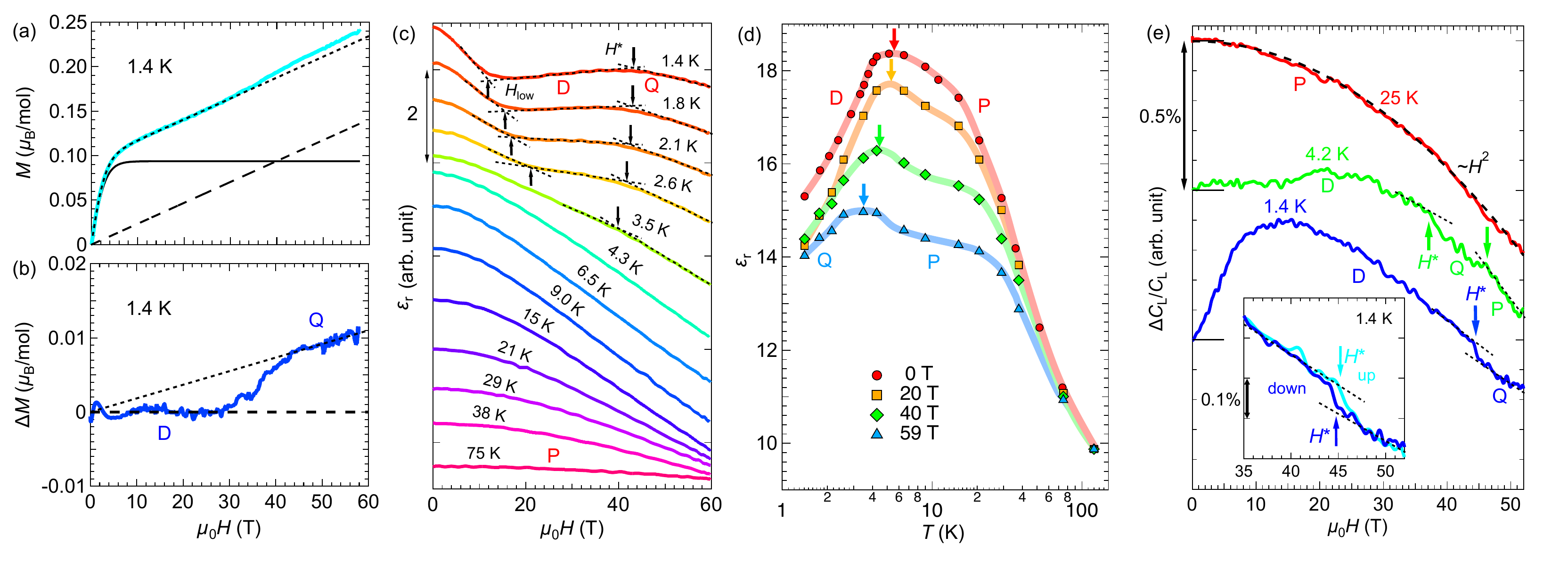}
\caption{
(a) High-field magnetization at 1.4~K.
Dotted curve is a fit of lower-field data ($<$30~T) based on a simple addition of the Brillouin function (solid line) and the linear term (dashed line).
(b) Residual magnetization obtained by subtracting the dotted curve shown in (a).
(c) Magnetic field dependence of dielectric permittivity $\varepsilon$$_{\rm r}$ at various temperatures.
Arrows indicate the anomalies at $H_{\rm low}$ of 10-20 T and $H^{\ast}$ of approximately 45~T.
(d) Temperature dependence of $\varepsilon$$_{\rm r}$ at $\mu_0$$H$ = 0, 20, 40, 59~T deduced from the field-dependent $\varepsilon$$_{\rm r}$ shown in (c).
Translucent curves behind data points serve as visual guides.
(e) Relative change in elastic constant $\Delta$$C_{\rm L}$/$C_{\rm L}$ as a function of field.
Dashed black curve indicates $H^2$ behavior.
At lower temperatures, $\Delta$$C_{\rm L}$/$C_{\rm L}$ shows anomalies related to transitions to the Q and P states, as indicated by the arrows.
Inset shows the 1.4~K data near $H^{\ast}$ in up (light blue) and down (blue) field sweeps.
}
\label{fig2}
\end{figure*}
As reported in Ref.~\cite{34}, magnetization below 30~T can be described by the simple summation of two components: the non-interacting Curie-type component (solid curve) and the linear term (dashed line).
The fitting of the data below 30~T reveals that the contribution of the former term amounts to 9.4$\pm$0.1$\%$ of the S = 1/2 Brillouin function at 1.4~K, decreasing to 7.3$\pm$0.1$\%$ at 4.2~K.
Notably, this concentration of spin solitons is anomalously large compared to typical values in conventional 1D organics (10$^{-5}$-10$^{-4}$\cite{11,11p3,11p5}).
Given the good crystallization of organic crystals, it becomes challenging to attribute such a significant quantity of spins to extrinsic impurities.
The amount of extrinsic impurities does not show any temperature dependence, whereas the former term in the present salt exhibits a strong dependence on temperature, indicating that it originates from the intrinsic spin solitons.
Magnetic torque measurements also indicates that this component cannot originate from extrinsic impurities (see Supplementary Materials\cite{Supp}).
Considering the proximity to the quantum critical point, the presence of dense solitons can be attributed to strong quantum fluctuations.
This is because dimerization fluctuations locally induce the formation of solitons\cite{34}.
In conventional SP systems, dilute solitons in the low-field state are regarded as a gas.
Even in the high-field SL state, the concentration of spin solitons immediately above $H^{\ast}$ is only 1$\%$-2$\%$ \cite{13,17}.
This fact indicates that the dense solitons in TTF-QBr$_3$I experience strong intersoliton interactions, which must lead to the solidification of the solitons.
Consequently, it is expected that the dense solitons no longer behave as a gas; however, the solitons retain their mobility due to the presence of strong quantum fluctuations, as reported in Ref.~\cite{34}.
Hence, the solitons should be a quantum liquid even in the low-field state.

Figure~\ref{fig2}a shows that the magnetization curve deviates slightly from the fit above 40~T.
For a clearer visualization, we present $\Delta$$M$ obtained by subtracting the abovementioned two components in Fig.~\ref{fig2}b.
The step-like increase in magnetization is consistent with the magnetization process near the transition to the SL state in conventional SP systems\cite{12,14,24,26}.
As will be discussed later, this change in magnetization demonstrates a transition to the Q state.

To examine the effect of magnetic field on ferroelectricity, the magnetic field dependence of dielectric constant $\varepsilon$$_{\rm r}$ at various temperatures is shown in Fig.~\ref{fig2}c.
The field-dependent dielectric response indicates the presence of magnetoelectric coupling, which facilitates cooperation between the ferroelectric and SP transitions through the dimerization, as in the case of TTF-BA\cite{31}.
Even at temperatures much higher than the zero-field $T_{\rm FSP}$, $\varepsilon$$_{\rm r}$ exhibits appreciable field dependence, which indicates that fluctuating dimerization appears much above $T_{\rm FSP}$.
Because of the relatively larger gap of TTF-QBr$_3$I, $\Delta$$_0$/$k_{\rm B}$=50-60 K\cite{34}, the short-range dimerization at temperatures much higher than $T_{\rm FSP}$ is reasonable.
At low temperatures, the field-dependent behavior exhibits two characteristic kinks at $H_{\rm low}$ and $H^{\ast}$.
As temperature increases, the lower-field anomaly at $H_{\rm low}$ broadens, and $H_{\rm low}$ shifts to higher fields.
The green triangles in Fig.~\ref{fig1}c show the temperature dependence of $H_{\rm low}$, which exhibits linear behavior.
These characteristics are reminiscent of the Brillouin function, which describes the magnetization process of spin solitons shown in Fig.~\ref{fig2}a.
Quantitatively discussing the relationship between $\varepsilon$$_{\rm r}$($H$) and the Brillouin function is challenging, as the magnetoelectric coupling cannot be estimated.
Nevertheless, the behavior suggests that the dielectric property is influenced by the magnetic polarization of the spin solitons.

In Fig.~\ref{fig2}c, another anomaly is observed at a higher field at $H^{\ast}$$\approx$45~T.
As in the case of magnetization (Fig.~\ref{fig2}b), this anomaly reflects a transition to the Q phase.
The temperature dependence of $H^{\ast}$ is not significant; however, the anomaly at $H^{\ast}$ vanishes gradually at elevated temperatures.
To view these results from a different perspective, we replot the datasets as the temperature dependence of $\varepsilon$$_{\rm r}$ in various fields in Fig.~\ref{fig2}d.
At 0 T, the D state appears below $T_{\rm FSP}$$\approx$5.6~K, which is determined by the peak temperature.
As the field increases, $T_{\rm FSP}$ decreases; however, the anomaly persists up to 59~T, as indicated by the arrows.
Based on the phase diagram shown in Fig.~\ref{fig1}c, the anomaly observed upon cooling under a field above $H^{\ast}$ must correspond to the transition to the Q phase, not the D phase.

As the FSP order originates from the molecular dimerization, the field-induced transition from the D to Q phases should be accompanied by changes in the elastic properties.
Figure~\ref{fig2}e presents the relative change in the elastic constant $\Delta$$C_{\rm L}$/$C_{\rm L}$.
At 25~K, where the state is P, as the magnetic field increases, $\Delta$$C_{\rm L}$/$C_{\rm L}$ decreases, namely the lattice softens.
This $H^2$-dependence is a background term, which is understood by the exchange-striction model\cite{36,37}.
At 1.4~K, $\Delta$$C_{\rm L}$/$C_{\rm L}$ increases with the magnetic field, i.e., lattice hardening is observed at low fields.
As in the case of $\varepsilon$$_{\rm r}$($H$) below $H_{\rm low}$, the low-field anomalous behavior is reminiscent of the Brillouin function.
In fact, the lattice hardening below $H_{\rm low}$ is consistent with the decrease in $\varepsilon$$_{\rm r}$($H$) in terms of the free energy change, and thus originates from the magnetization process of the spin solitons. 
In conventional SP state, such hardening is not observed\cite{28} because dilute solitons cannot affect the bulk elastic properties.
This observation suggests that the dense solitons have a substantial impact on the bulk elastic properties.
At $H^{\ast}$, as in the case of $M$ and $\varepsilon$$_{\rm r}$, $\Delta$$C_{\rm L}$/$C_{\rm L}$ shows an anomaly related to the transition to the Q phase, as indicated by the arrows.
The inset is an enlarged plot of the 1.4~K data around $H^{\ast}$, which indicates that the anomaly at $H^{\ast}$ shows hysteresis depending on the field-sweep direction.
The transition at $H^{\ast}$ is a first-order transition, sharing a similarity with the first-order SP-SL transition of the conventional SP systems\cite{26}.
It is worth emphasizing that the transition between the D and Q states results in only the slight softening.

\begin{figure}
\centering\includegraphics[width=0.7\hsize]{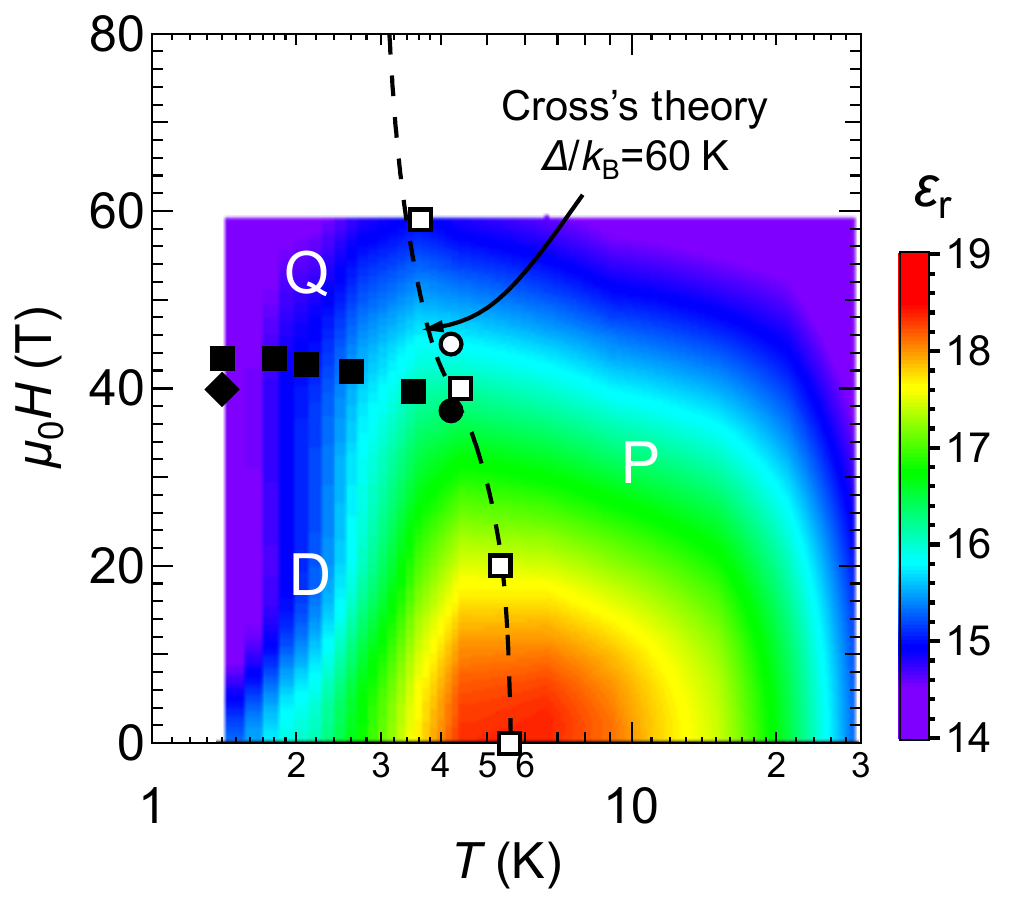}
\caption{
Semilogarithmic plot of $H$-$T$ phase diagram of the FSP state with a color plot of $\varepsilon$$_{\rm r}$.
The definition of the symbols is the same as those in Fig.~\ref{fig1}c.
Dashed curve shows the phase boundary expected in Cross's theory\cite{19}.
}
\label{fig3}
\end{figure}

Based on these results, we construct the field-temperature phase diagram of the FSP state with the color plot of $\varepsilon$$_{\rm r}$, as shown in Fig.~\ref{fig3}. 
The color variation highlights the phase boundary between the P and FSP states.
The open and filled squares, which determine the phase boundaries, are obtained from the temperature and field dependences of $\varepsilon$$_{\rm r}$, respectively.
The circles at 4.2~K signify the fields of anomalies in $\Delta$$C_{\rm L}$/$C_{\rm L}$.
Here, we compare the high-field Q phase with the SL state in conventional SP systems.
The shape of the present phase diagram is similar to that of the conventional SP systems shown in Fig.~\ref{fig1}a.
Nevertheless, the transition field $\mu_0$$H^{\ast}$$\approx$45~T appears much higher than that expected from the ratio $\mu_0$$H^{\ast}$/$T_{\rm SP}$ = 0.9-1.2~(T/K) for conventional SP states\cite{12,14,19,26}.
Robustness against magnetic fields is observed in TTF-BA as well\cite{31}.
These TTF-based salts show a short-range FSP correlation via fluctuating dimerization at temperatures much higher than $T_{\rm FSP}$ and exhibits a large $\Delta$$_0$/$k_{\rm B}$\cite{31,34}.
The strong quantum fluctuations in TTF-QBr$_3$I suppress the long-range ordering temperature $T_{\rm FSP}$ significantly; thus, the relationship between $T_{\rm FSP}$ and $\mu_0$$H^{\ast}$ is no longer valid.
Because the stability of the SP state is governed by the balance between the Zeeman energy and spin gap, $\Delta$$_0$/$k_{\rm B}$ instead of $T_{\rm SP}$ is more suitable for evaluating $\mu_0$$H^{\ast}$.
For conventional SP states\cite{6,24,27,38}, $\mu_0$$H^{\ast}$/($\Delta$$_0$/$k_{\rm B}$) is typically in the range of 0.5-0.7~(T/K).
Therefore, $\Delta$$_0$/$k_{\rm B}$ = 50-60~K for TTF-QBr$_3$I\cite{34} indicates that the order of the obtained $\mu_0$$H^{\ast}$ = 40-45~T is reasonable.
Cross's theory\cite{19} proposes the universal phase boundary of the SP state using a $\mu_{\rm B}$$H$/(4$\pi$$k_{\rm B}$$T_{\rm SP}$(0)) vs. $T_{\rm SP}$($H$)/$T_{\rm SP}$(0) plot, which agrees well with the experimental results of conventional SP systems\cite{12,26}.
Using the relationship $\Delta$$_0$$/$$k_{\rm B}$ = 1.7$T_{\rm SP}$(0)\cite{6,24,27,38}, we plot the phase boundary of Cross's theory for the case of $\Delta$$_0$/$k_{\rm B}$ = 60~K (as shown in Fig.~\ref{fig3}) and discover that the curve reproduces the present results.
Although the FSP state shows ferroelectric polarization, the agreements indicate that the magnetic-field response of the FSP state is governed by the Zeeman effect and that it can be understood based on the framework of Cross's theory.

Next, we consider the difference between the conventional SP and present FSP states.
In the case of conventional SP systems, the solitons in the D state are dilute\cite{11} and, therefore, should behave as an almost non-interacting gas. 
In high-field SL states, solitons emerge at the domain walls and form a superlattice with periodic potential.
The formation of the lattice corresponds to the solidification of the solitons through interactions between the solitons.
The gas--solid transition induces strong lattice hardening along the modulation direction, as reported in the previous studies\cite{28,30}.
By contrast, for the present FSP state, the transition to the Q phase shows only a slight lattice softening at $H^{\ast}$, as shown in Fig.~\ref{fig2}e.
This suggests that the solitons in the Q state are not able to form a lattice.
The solitons in the present D state should be a quantum liquid.
Hence, this change at $H^{\ast}$ is attributable to a liquid--liquid transition, which does not show significant changes owing to the absence of symmetry breaking.
In fact, the field dependence of $\varepsilon$$_{\rm r}$ shows only a slight decrease in $\varepsilon$$_{\rm r}$ at $H^{\ast}$.
If the solitons formed a rigid lattice, their mobility would siginificantly diminish, leading to a sharp decrease in $\varepsilon$$_{\rm r}$ at $H^{\ast}$ because $\varepsilon$$_{\rm r}$ reflects the soliton dynamics.
Therefore, the slight disparity in $\varepsilon$$_{\rm r}$ between the D and Q states also indicates that the correlated solitons maintain their mobility even in the Q state.
This finding stands in stark contrast to the long-range soliton ordering observed in the conventional SP systems\cite{17,35p7}.

The question arises: why do the solitons exhibit liquid-like behavior even in the Q state?
Let us first consider an extrinsic factor.
The presence of disorder is expected to impede the long-range ordering of solitons.
However, in such scenarios, solitons become localized due to pinning at the minima of a static random potential, leading to the absence of their liquidity at low temperatures.
Next, we delve into an intrinsic factor unique to the present FSP state.
The present system is influenced by quantum fluctuations\cite{34}; thus, quantum melting must be considered.
When the intersoliton distance $L$ becomes comparable to the amplitude of the fluctuations in the soliton position, quantum melting occurs.
As the magnetization of the spin solitons at 50~T is approximately 10$\%$ of the full moment, the average $L$ reaches approximately 10$d$, where $d$ represents the intersite distance (half the $b$-axis length).
We reiterate that for conventional SP systems, the concentration of spin solitons is only 1$\%$-2$\%$ immediately above $H^{\ast}$\cite{13,17}; thus, $L$ is 50$d$-100$d$.
In the classical model for conventional SP states, the soliton width $\xi$ is expressed as $\xi$/$d$ = $\pi$$J$/2$\Delta$$_0$ in the SL state\cite{7,13,20,21}; thus, $\xi$/$d$ of TTF-QBr$_3$I is estimated to be 3-3.5.
If the effect of quantum fluctuations are further incorporated, the soliton width is expected to be larger\cite{41,42}, and the actual $\xi$/$d$ value must be higher than the calculated value.
In fact, in TTF-QBr$_3$I, the quantum fluctuations render the effective mass of the solitons several hundred times lighter than the masses of the molecules\cite{34}, which is attributable to the augmentation of the soliton width\cite{41,42,43}.
If the enhanced $\xi$ reaches a length comparable to $L$, then the SL is replaced with the Q state, which is regarded as a quantum liquid state due to the quantum melting of the SL.
Since the results shown in Fig.~\ref{fig2} indicate that the difference between the D and Q states is insignificant, the transition between these states only slightly modify the intersoliton interactions and the density of solitons. 
Our results indicate that the transition between the D and Q states is a quantum liquid--quantum liquid transition of the spin solitons induced by a magnetic field.

In the present study, we discovered that the FSP state in TTF-QBr$_3$I shows the dense spin solitons even at low fields because of the strong quantum fluctuations, and that the solitons behave as a quantum liquid.
The FSP state exhibits a transition to the high-field Q phase.
Even in the Q phase, strong quantum fluctuations inhibit the formation of the superlattice of the solitons, and the spin solitons remain as a quantum liquid.
This fact indicates the magnetic-field-induced transition of topological particles from a quantum liquid to another quantum liquid.

This study was partly supported by JSPS KAKENHI Grant (20K14406, 22H04466) and JST CREST Grant (JPMJCR18J2).

\renewcommand{\thefigure}{S\arabic{figure}}
\clearpage
\onecolumngrid
\appendix
\begin{center}
\large{\bf{Supplementary Materials for\\
Quantum liquid states of spin solitons in a ferroelectric spin-Peierls state
}}
\end{center}

\section{Experimental details}
\begin{figure}[hh]
\begin{center}
\includegraphics[width=0.5\linewidth,clip]{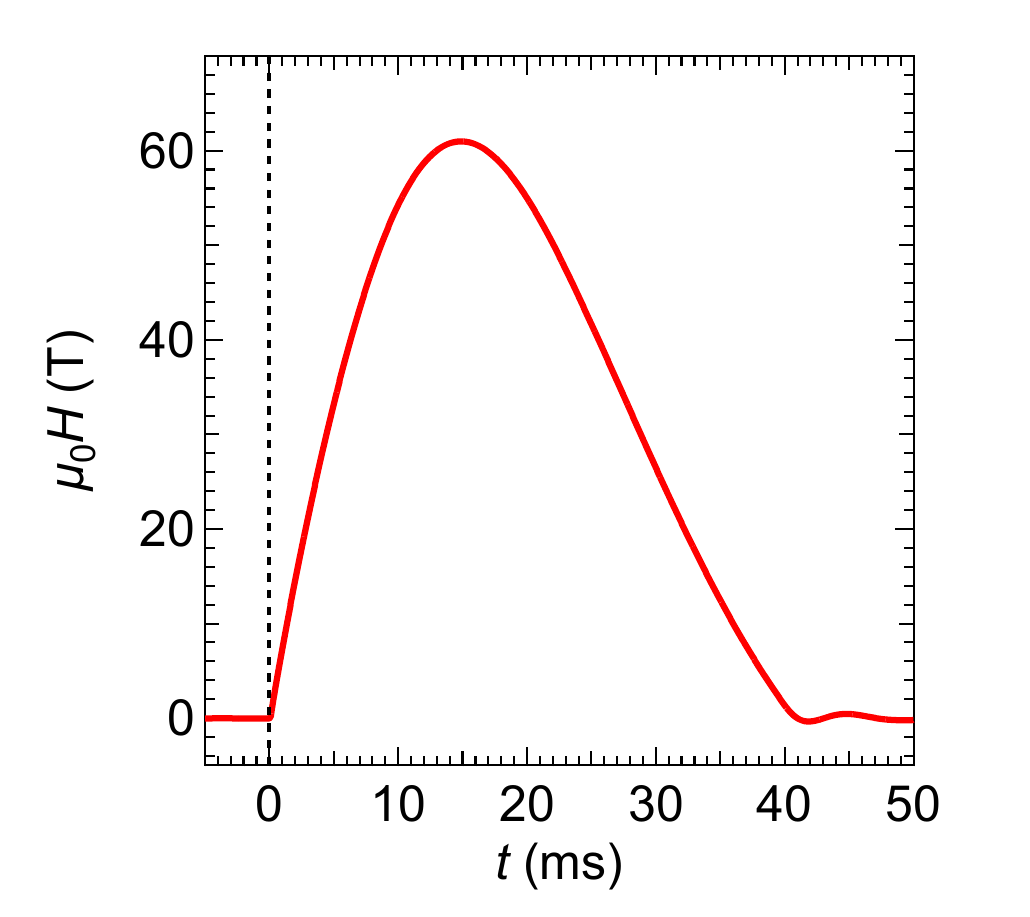}
\end{center}
\caption{
Typical time-field profile of pulsed magnetic field used to measure various physical properties in this study.
}
\label{figS1}
\end{figure}

Single crystals of TTF-QBr$_3$I were grown by slow evaporation of a cold mixed acetonitrile solution comprising TTF and QBr$_3$I.
Magnetization curves were obtained for the polycrystalline samples.
The relative dielectric permittivity $\varepsilon$$_{\rm r}$ was measured using 100~kHz AC electric fields applied along the $b$ axis, which is parallel to the molecular columns\cite{35p5}.
For the $\varepsilon$$_{\rm r}$ and $C_{\rm L}$ measurements, magnetic fields were applied along the $b$ axis.
The elastic constant $C_{\rm L}$ was measured using 32.0~MHz longitudinal sound waves along the $b$ axis.
The magnetic fields in these measurements were generated using an in-house 60~T pulse magnet.
Figure~\ref{figS1} shows the typical time-field profile of the pulsed magnetic field generated in this study.

\newpage
\section{Detailed discussion of the origin of the Curie-type component}
\begin{figure}[hh]
\begin{center}
\includegraphics[width=0.8\linewidth,clip]{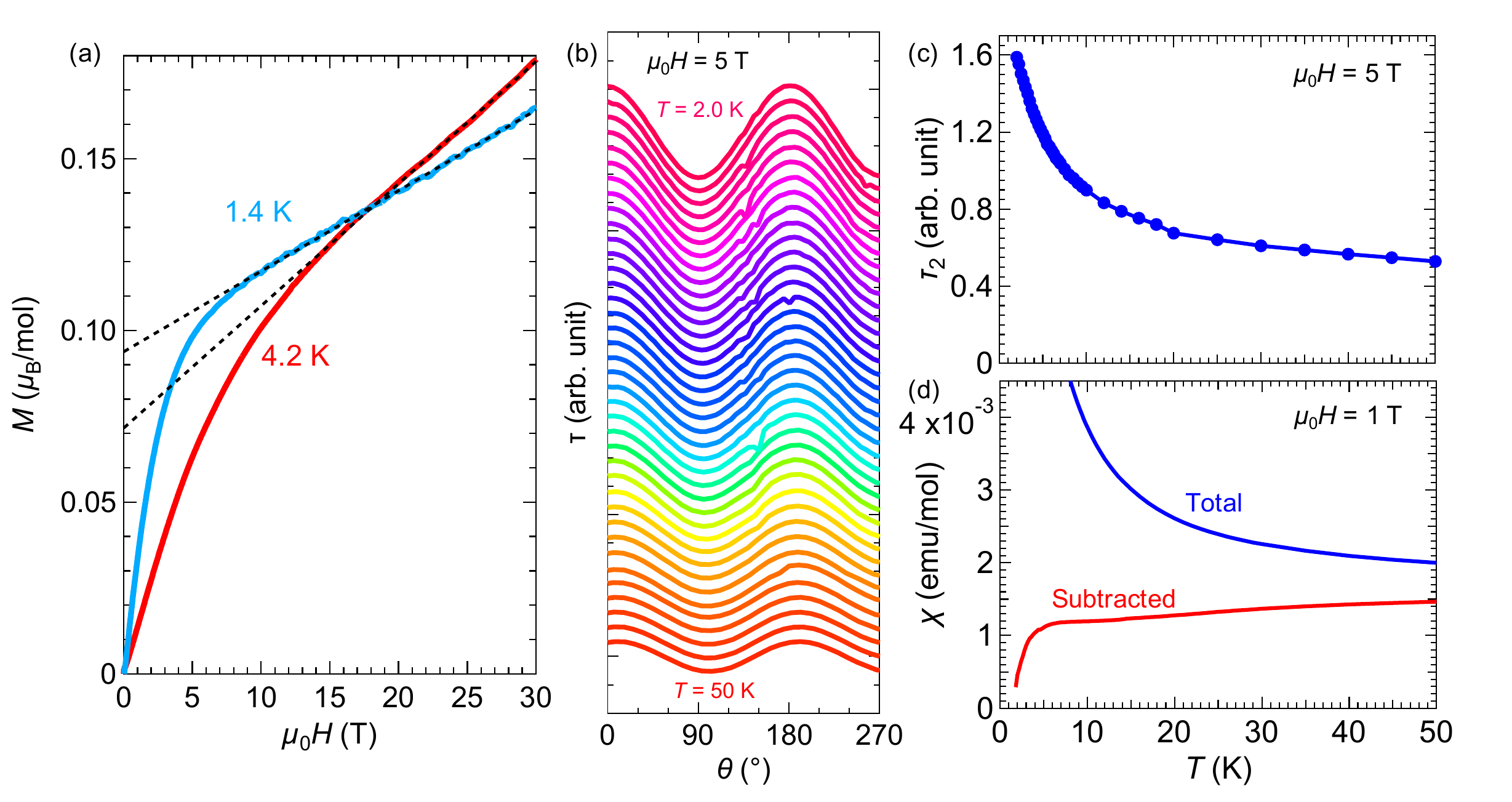}
\end{center}
\caption{
(a) Magnetization curves at 1.4~K and 4.2 K.
Dotted curves indicates linear fits to the data in the range of 20-30~T.
(b) Angle-dependent magnetic torque of TTF-QBr$_3$I at various temperatures under a field of 5~T.
When the angle $\theta$ is 0$^{\circ}$, the direction of the applied magnetic field is parallel to the $c^{\ast}$ axis.
(c) Temperature dependence of $\tau$$_2$ obtained from the data shown in (a).
(d) Temperature dependence of magnetic susceptibility $\chi$ of TTF-QBr$_3$I (blue curve). 
Red curve represents the component of antiferromagnetically interacting spins in a one-dimensional chain, obtained by subtracting the Curie term\cite{34}.
}
\label{figS2}
\end{figure}

In the main text, we provided a discussion on the origin of the Curie-type component observed in low-temperature magnetization measurements.
Our analysis suggests that the predominant source of the Curie component is the presence of spin solitons.
However, in order to arrive at this conclusion, it is necessary to eliminate the effect of the presence of isolated spins originating from extrinsic impurities or disorder, given thatTTF-QBr$_3$I inherently exhibits disorder due to the occupational disorder of the halogen sites\cite{35}.
Here we discuss the origin of the Curie component in further detail.
As discussed in the main text, the contribution of the Curie term amounts to 9.4$\pm$0.1$\%$ at 1.4~K, decreasing to 7.3$\pm$0.1$\%$ at 4.2~K.
These values are derived from fitting data below 30~T, employing a function combining the Brillouin function and a linear term.
While it is acknowledged that spin solitons may experience a weak internal magnetic field from surrounding antiferromagnetically interacting spins, the Brillouin-function-like behavior saturates above 20~T at such low temperatures.
Consequently, the magnitude of the Curie term can be estimated from the intercept obtained through a linear fit of data within the 20-30 T range, as shown in Fig.~\ref{figS2}(a).
The linear fit also corroborates that the Curie term amounts to 9.4$\%$ at 1.4~K and 7.2$\%$ at 4.2~K, with errors of $<$0.1$\%$.
Since the amount of extrinsic impurity and structural disorder naturally show no temperature dependence, the significant temperature dependence strongly indicates that the Curie term originates from the intrinsic spin solitons.

As shown in Fig.~2, the bulk elastic properties detect the magnetization process of the spin solitons.
As reported in Ref.~\cite{35}, x-ray diffraction data reveals an $R$-factor of approximately 0.03, signifying good crystallization and uniform iodine atom distribution across all sites, each comprising 20-30$\%$.
This observation implies an absence of domain structure, establishing the crystal as an average structure with the local occupational disorder.
Should such localized disorder instigate the observed spin solitons, each soliton is confined within an individual molecule, with the soliton's effective mass corresponding to that of TTF/QBr$_3$I molecules.
However, it is noteworthy that the soliton mass is several hundred times smaller than that of the molecules and exhibits a broad soliton width\cite{34}.
This discrepancy negates the possibility of spin solitons being induced solely by the occupational disorder.

Furthermore, angle-dependent magnetic torque measurements were conducted at varying temperatures, as depicted in Fig.~\ref{figS2}(b).
The magnetic torque $\tau$ is governed by the relation $\tau$ = $M$ $\times$ $H$, leading to an angular dependence $\tau$($\theta$) expressed as proportional to ($M_x$$-$$M_y$)$H$sin(2$\theta$) = $\tau$$_2$sin(2$\theta$) in a paramagnet.
Here, $M_x$ and $M_y$ denote magnetization along the respective principal axes, and the amplitude of the twofold term $\tau$$_2$ correlates with the anisotropy of magnetization.
Impurity spins, less susceptible to surrounding molecular interactions, exhibit no anisotropy and, consequently, do not manifest in magnetic torque measurements.
Indeed, magnetic torque measurements are commonly employed to eliminate extrinsic impurities such as the Curie tail, isolating intrinsic low-temperature magnetic components\cite{S1,S2,S3}.
In Fig.~\ref{figS2}(c), we present the temperature-dependent behavior of $\tau$$_2$. 
For comparison, the temperature dependence of static magnetic susceptibility, measured by SQUID magnetometry\cite{34}, is shown in Fig.~\ref{figS2}(d).
To confirm the low-temperature behavior, the magnetic susceptibility after subtracting the Curie term is also shown as the red curve.
The low-temperature behavior of $\tau$$_2$ is qualitatively similar to that of the non-subtracted magnetic susceptibility.
While the precise behavior is challenging to evaluate due to the temperature-dependent Curie term, $\tau$$_2$ evidently includes the Curie term.
Consequently, the origin of the Curie term cannot be ascribed to extrinsic non-interacting impurities and/or disorder.
Given that spin solitons exhibit mobility owing to the repeated creation and annihilation of dimerization in the one-dimensional chain, they are detectable by magnetic torque measurements because of the inherent anisotropy.

\end{document}